\documentclass[twocolumn,prb,showpacs]{revtex4}
\usepackage{graphicx}
\usepackage{amssymb}
\usepackage{bm}

\newcommand{\GTO}{$\rm Gd_2Ti_2O_7$}
\newcommand{\GSO}{$\rm Gd_2Sn_2O_7$}
\begin{document}

\title{Magnetic resonance in a pyrochlore antiferromagnet
Gd$_2$Ti$_2$O$_7$}

\author{S. S. Sosin, A. I. Smirnov, L. A. Prozorova}
\affiliation{P. L. Kapitza Institute for Physical Problems RAS,
119334 Moscow, Russia}
\author{G. Balakrishnan}
\affiliation{Department of Physics, University of Warwick,
Coventry CV4 7AL, UK}
\author{M. E. Zhitomirsky}
\affiliation{Commissariat \`{a} l'Energie Atomique,
DSM/DRFMC/SPSMS, 38054 Grenoble, France}

\date{\today}

\begin{abstract}
Electron spin resonance study of frustrated pyrochlore
Gd$_2$Ti$_2$O$_7$ is performed in a wide frequency band for a
temperature range 0.4--30~K, which covers paramagnetic and
magnetically ordered phases. The paramagnetic resonance reveals
the spectroscopic $g$-factor about 2.0 and a temperature dependent
linewidth. In ordered phases magnetic resonance spectra are
distinctive for a nonplanar cubic (or tetrahedral) antiferromagnet
with an isotropic susceptibility. In the high-field saturated
phase, weakly-dispersive soft modes are observed and their field
evolution is traced.
\end{abstract}


\maketitle

Frustrated antiferromagnets are among the most interesting
strongly correlated magnetic systems. Geometry of a crystal
lattice in these compounds does not allow the energy of all paired
exchange bonds to be minimized simultaneously. In the case of
pyrochlore geometry of corner-sharing tetrahedra frustration leads
to a macroscopic number of classical ground states. Strong
fluctuations between degenerate ground states preclude any
conventional type of magnetic ordering. \cite{villain} Eventually,
degeneracy is removed by weak residual interactions, (dipolar,
next-nearest-neighbor exchange, relativistic anisotropy) or by
short wave-length fluctuations (order by disorder effect). Such a
degeneracy lifting occurs with the two major consequences: (i) if
a long-range ordering does occur in such a system, it develops at
temperatures much lower than in a usual nonfrustrated
antiferromagnet with a similar magnitude of the exchange integral
(ii) a delicate balance between a number of weak interactions
yields a large variety of possible outcomes so that similar
compounds can reveal drastically different magnetic properties.

A nice example of the above principles is provided by two
pyrochlore compounds \GTO\ and \GSO. Magnetic Gd$^{3+}$ ions with
semiclassical spins $S=7/2$ have zero orbital momentum and,
consequently, reduced crystal field effects compared to other
rare-earth compounds. Both gadolinium pyrochlores have complicated
phase diagrams. Magnetic ordering in \GTO\ takes place at
$T_{c1}\simeq 1$ K followed by a second transition at
$T_{c2}\simeq 0.75$ K. \cite{ramirez,petrenko} At $T<T_{c2}$ two
field-induced transitions are observed under magnetic field at
$H_{c1}\simeq 30$ kOe and $H_{c2}\simeq 60-70$~kOe (above $H_{c2}$
the magnetic moment is close to saturation). A single first-order
phase transition occurs in \GSO\ at $T_c\simeq 1$ K.
\cite{bonville} Recent neutron diffraction study have identified
the low temperature phase of \GTO\ to be a multi-$k$ structure
described by four wave-vectors  of the type ${\bf
k}=(\frac{1}{2}\,\frac{1}{2}\,\frac{1}{2})$ with partial disorder
in the intermediate phase. \cite{stewart} This type of structure
can be stabilized due to dipolar interaction in combination with a
next-nearest-neighbor superexchange.\cite{raju,cepas,wills}
Another possibility is provided by strong fluctuations choosing
the state with ${\bf k}=(0\,0\,0)$, \cite{palmer,cepas} which was
recently observed in \GSO.

An important consequence of a macroscopic number of nearly
degenerate states in  a geometrically frustrated magnet is that
its excitation spectrum contains soft, almost dispersionless modes
in the whole Brillouin zone. Neutron experiments \cite{lee} in the
frustrated spinel compound ZnCr$_2$O$_4$ indicate such modes to
exist above the ordering temperature. Almost degenerate modes lead
to a large residual entropy and an enhanced magnetocaloric effect,
\cite{misha} which has been measured in \GTO\ and a garnet
Gd$_3$Ga$_5$O$_{12}$. \cite{sosin} The dynamics of these modes
below ordering cannot be treated as a conventional
antiferromagnetic resonance. Uniform oscillations of spins in
crystal near an equilibrium state (fixed by weak forces) are
accompanied by an absence of spin stiffness usually provided by
the strongest exchange interaction. It is not obvious {\it a
priori}, whether such modes should be resonance spin oscillations
or a kind of relaxation movement. Since inelastic neutron
scattering experiments are hindered by a large nuclear
cross-section of gadolinium ions, the ESR technique remains the
only possibility to probe the low-energy spin dynamics  in \GTO
and \GSO.

In the present work we study a single crystal of \GTO\ in a wide
frequency and field ranges at temperatures both above and below
$T_{c1,2}$. The magnetic resonance signals are observed and
analyzed in the paramagnetic and in all four ordered phases. At
the lowest experimental temperature 0.42 K two resonance branches
with different energy gaps are observed at fields below $H_{c1}$.
In the intermediate range $H_{c1}<H<H_{c2}$ the field-dependent
gap of the magnon branch, softening in the vicinity of the
spin-flip transition $H_{c2}$ is found. In the spin-saturated
phase three weak and narrow resonance modes are detected, two of
them corresponding to ``soft modes'' with weak dispersion in
agreement with the spin-wave calculations.

\begin{figure}[t]
\centerline{\includegraphics[width=\columnwidth]{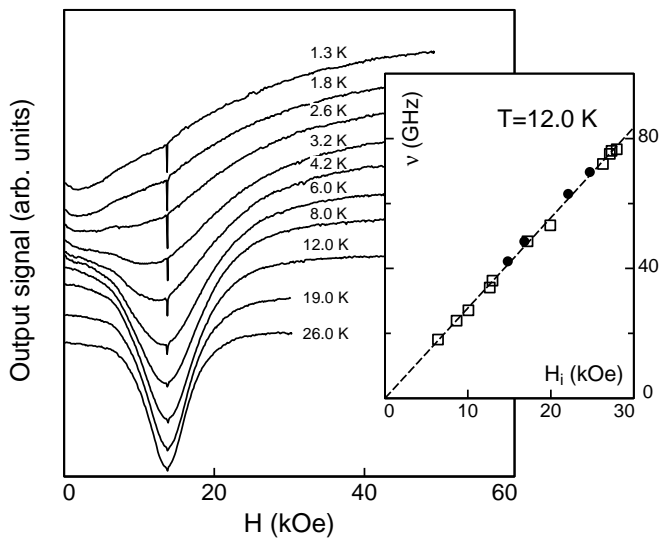}}
\caption{The field dependence of the resonance absorption in \GTO\
at $\nu=36.2$ GHz ($H\perp [111]$ axis) for various temperatures
above $T_{N1}$. The inset shows the frequency-field diagram for
$T=12$ K (closed and open circles correspond to $H\perp [111]$ and
$H\parallel [111]$ respectively); dashed line is a paramagnetic
resonance with $g=2$.} \label{fig1}
\end{figure}

A single crystal of \GTO\ has been grown by a floating-zone
technique. \cite{oleg} We studied two small samples cut in a form
of thin square plates of approximately $1\times 1\times 0.1$ mm in
size (0.5--0.6 mg by mass) with their $[111]$-axes oriented
parallel and perpendicular to the sample plane. The resonance
measurements are performed in a set of transmission type
spectrometers with different resonators covering the frequency
range 20--100~GHz. An external magnetic field up to 120 kOe
created by a cryomagnet was always applied in a sample plane so
that no correction for the demagnetization factor is required.

Measurements in the temperature range 1.2--30 K are carried out in
a $^4$He cryostat. The corresponding field scans taken at a
frequency 36.2 GHz for various temperatures are shown in fig.
\ref{fig1}. At high temperatures $T\gg\theta_{CW}\simeq 10$~K a
single resonance line with the linewidth $\Delta H\sim 3$ kOe is
observed in all magnetic field orientations. It has a linear
frequency-field dependence with an isotropic $g$-factor 2.0 (see
inset in Fig.~\ref{fig1}). On decreasing the temperature below 10
K the resonance line starts to broaden and decrease in the
amplitude. In the vicinity of the upper ordering transition
$T_{c1}$ the line transforms into a wide nonresonant band of
absorption having its maximum at zero magnetic field. The
splitting of the resonance line at high frequencies previously
reported in Ref.~\onlinecite{hassan} and attributed to the strong
anisotropic effects has been also observed in our experiment. We
have, however, found that such a splitting exists only for large
samples and is easily eliminated by decreasing the size of the
plate. This effect should therefore be attributed to the
electrodynamic resonances in the dielectric sample with the field-
and temperature-dependent magnetic susceptibility. Detailed
analysis of the high temperature resonance properties of \GTO\ and
\GSO\ will be published elsewhere.

\begin{figure}[b]
\centerline{\includegraphics[width=\columnwidth]{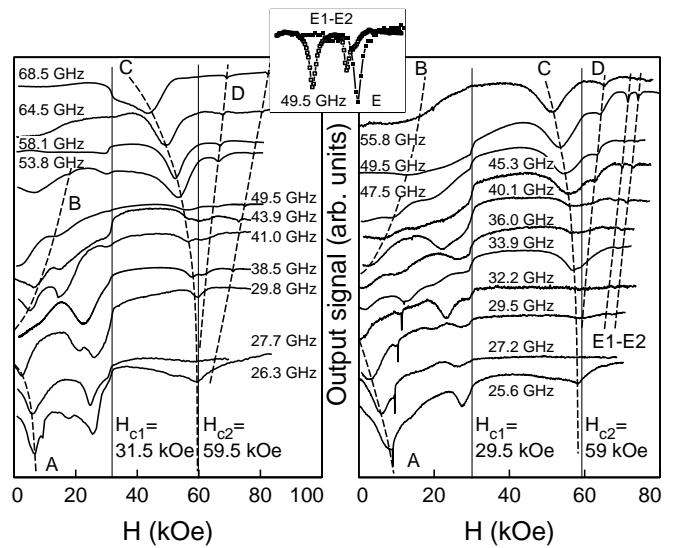}}
\caption{The absorption spectra recorded at $T=0.42$ K at
different frequencies (increasing bottom-up) for $H\parallel
[111]$ (left panel) and $H\perp [111]$ (right panel). Dashed lines
are guide-to-eyes to trace the frequency evolution of the same
resonance modes, vertical solid lines mark the points of the phase
transitions $H_{c1}$ and $H_{c2}$. The inset expands the weak
resonance lines E and E1-E2 taken at the same frequency in both
field orientations.} \label{fig2}
\end{figure}

The low-temperature ESR spectra have been obtained in a $^3$He
cryostat equipped with a sorb pump enabling to cool down to
approximately 0.4 K. The resonance absorption spectra taken at the
lowest temperature 0.42 K in the whole frequency range are
presented in Fig.~2 for two principal orientations of the external
field ($H\parallel [111]$ and $H\perp [111]$ left and right panels
respectively). Among a number of resonance lines observed below
$H_{c1}$ in both orientations one can trace absorptions belonging
to two different gapped branches, one of them decreasing (line A)
and the other one increasing (line B) in field (shown by dashed
guide-to-eyes in the fig. \ref{fig2}). Other resonances are
difficult to trace on changing the frequency, but all of them
soften or disappear in the vicinity of $H_{c1}$. This transition
is clearly observed in both field orientations by sharp jumps at
the absorption lines in both the forward and the backward field
sweeps without hysteresis. Just above the transition a new
resonance line appears at a frequency of about 70 GHz (line C). It
gradually drops on increasing the magnetic field practically
independent on its orientation and almost softens at the second
transition $H_{c2}$. This transition is accompanied by appearance
of two additional weak resonance modes (lines D and E) for
$H\parallel [111]$, both increasing in field. In a perpendicular
field orientation the line D remains unchanged while the line E is
splitted into two components (expanded on the inset in
Fig.~\ref{fig2}).

\begin{figure}
\centerline{\includegraphics[width=\columnwidth]{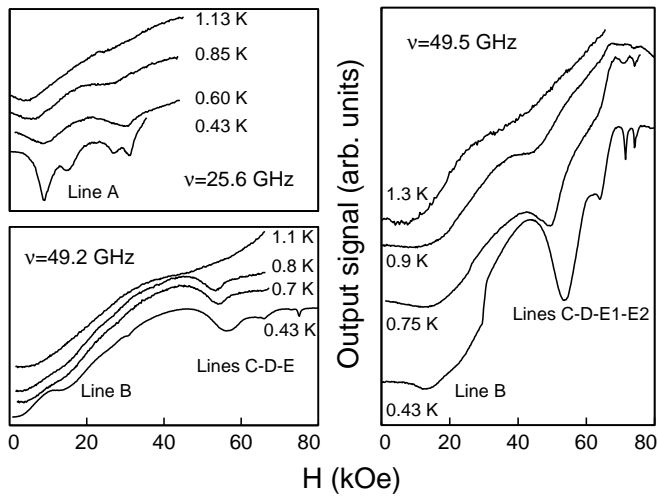}}
\caption{ The temperature evolution of the absorption lines at
given frequencies for $H\parallel [111]$ (left panel) and $H\perp
[111]$ (right panel).} \label{fig3}
\end{figure}

All resonance lines have a strong temperature dependence. On
heating they broaden and shift (see Fig.~\ref{fig3}) and finally
disappear while the background absorption line acquires the shape
observed in the preliminary high-temperature experiment. One
should mention that this nonresonant low-field background also
survives at temperatures below ordering.

Convenient approach to describe the low-field resonance properties
of magnetically ordered systems is a macroscopic hydrodynamic
theory, \cite{marchenko} which is applicable if an exchange
interaction plays the principal role in magnetic ordering. Such an
assumption might be incorrect for \GTO, a system with a
considerable single-ion anisotropy \cite{glazkov} and ordering
driven by weak interactions. We shall, however, use such a theory
in attempt to obtain a qualitative description of the experimental
data. A nonplanar exchange structure can be described by three
orthogonal unit vectors ${\bf l}_n$. \cite{marchenko} For the
$4k$-structure found in neutron experiments \cite{stewart} the
antiferromagnetic vectors ${\bf l}_n$ transform according to the
three-dimensional irreducible representation $F_1$ of the
tetrahedral group $T_d$. \cite{LL} The energy of magnetic
anisotropy is obtained by constructing various invariants from the
components $l^\alpha_n$. In the lowest order the anisotropy is
expressed as
\begin{eqnarray}
E_a & = & \lambda_1 \left(l_{1x}^2 + l_{2y}^2 + l_{3z}^2\right ) +
\lambda_2 \left (l_{1y} l_{2x} + l_{1z} l_{3x} + l_{2z} l_{3y}
\right ) \nonumber \\
 &  & \mbox{} +\lambda_3 \left (l_{1x} l_{2y} + l_{2y} l_{3z} + l_{3z}
l_{1x} \right ) \ , \label{ep}
\end{eqnarray}
where $\lambda_n$ are phenomenological parameters. The expression
(\ref{ep}) has five extremums, each can correspond to an
equilibrium orientation of ${\bf l}_n$ at some values of
$\lambda_n$. Three of these states preserve the original cubic
symmetry in which case the spectrum of uniform oscillations
consists of three degenerate eigenmodes while experimentally we
observe at least two different resonance branches at $H=0$. Among
two nonsymmetric states, the most probable one is obtained by
rotating the triad ${\bf l}_1={\bf x}$, ${\bf l}_2={\bf y}$, ${\bf
l}_3={\bf z}$ by an angle $\varphi =\arccos -
(2\lambda_1-\lambda_2+2\lambda_3)/[4(\lambda_1+\lambda_2+\lambda_3)]$
about the [111] direction.

The kinetic energy of small homogeneous oscillations of the
exchange structure in weak magnetic field $H\ll H_{c2}$ is
determined by susceptibility tensor, \cite{marchenko} which has an
isotropic form for the tetrahedral symmetry. (Magnetization
measurements at $T$=0.3 K \cite{Narumi} confirm that the values of
$\chi$ are equal for $H\parallel$ [111] and $H\parallel$ [110].)
The kinetic energy is written, therefore, as
\begin{equation}
K =\frac{\chi}{2} \left(\frac{\Omega}{\gamma}+ H\right)^2 \ ,
\label{ek}
\end{equation}
where $\Omega$ is an angular velocity, $\gamma =g\mu_B/\hbar$ is a
gyromagnetic ratio. Performing standard variation procedure, one
gets the following cubic equation for eigen frequencies:
\begin{equation}
(\nu^2-\Delta_1^2)(\nu^2-\Delta_2^2)^2 -
\tilde{\gamma}^2\nu^2(\nu^2H^2-\Delta_1^2H_{\parallel}^2
-\Delta_2^2H_{\perp}^2) =0 \ , \nonumber
\end{equation}
\noindent where $\Delta_{1,2}$ are the gap values (certain
combination of $\lambda_n$ and $\chi$ which can be treated as
fitting parameters), $H_{\parallel}$ and $H_{\perp}$ are field
components with respect to [111] axis,
$\tilde{\gamma}=\gamma/2\pi$.

\begin{figure}
\centerline{\includegraphics[width=8cm]{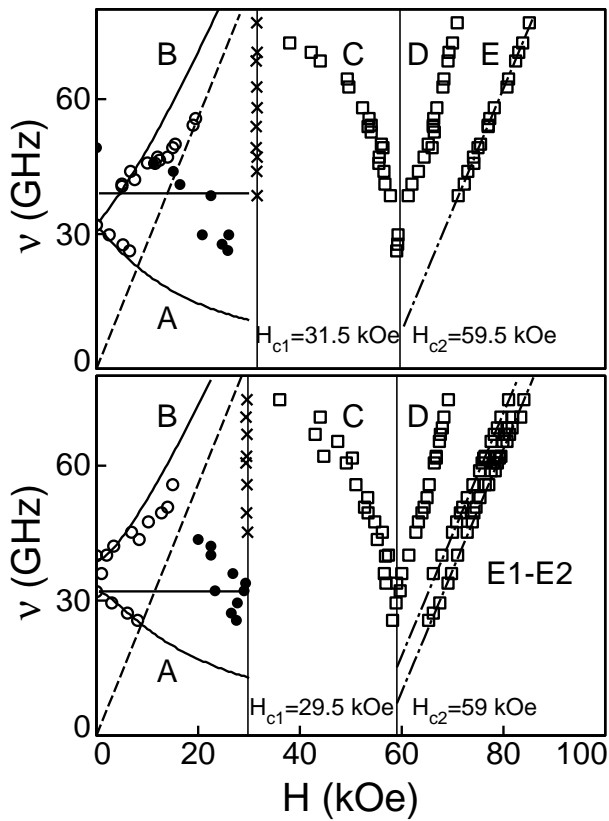}}
\caption{Frequency field diagram of the resonance spectrum in
\GTO\ at $T=0.42$ K for $H\parallel [111]$ (upper panel) and
$H\perp [111]$ (lower panel); {\Large $\circ$} -- low field lines,
$\Box$ -- high field lines, {\Large $\bullet$} -- unidentified
resonances, $\times$ -- $H_{c1}$ transitions; solid lines
represent the low field theoretical calculations with the
parameters $\Delta_1=39$ GHz, $\Delta_2=32$ GHz, dashed line is a
paramagnetic resonance with $g=2.0$, dashed-dotted lines are
linear fits as described in the text. Phase transitions at
$H=H_{c{1,2}}$ are pointed by vertical lines as in fig.
\protect\ref{fig2}} \label{fig4}
\end{figure}

The calculated field dependencies of all three branches for both
regular field orientations are shown in Fig.~\ref{fig4} by solid
lines with $\Delta_1=39$~GHz, $\Delta_2=32$~GHz. One can see that
the description of the spectrum in the range of theory application
(namely of the low field lines A and B) is satisfactory. However,
one should mention that this theory predicts all low-lying
excitations to have gaps, which excludes the experimentally
observed power-law dependence of the specific heat down to 0.2~K.
\cite{yaouanc} It also fails to explain the broad background
absorption observed in an ordered phase which might reflect the
non-frozen spin behavior at low temperatures. Consequently, an
``exchange softness'' of the magnetic structure in \GTO\ remains
essential down to temperatures $T\ll T_c$ and requires new
theoretical approaches for final description of the spin dynamics.

At high fields $H>H_{c2}$ the ground state simply corresponds to a
ferromagnetic alignment of spins. In the model with only the
nearest-neighbor exchange $J$ the four spin-wave modes (according
to a number of ions in a unit cell) are given as
\begin{equation}
\nu_{1,2}=\tilde{\gamma}H-8JS \ , \
\nu_{3,4}=\tilde{\gamma}H-2JS(2\mp \sqrt{1+3\eta_{\bf k}})
\label{softmodes}
\end{equation}
where $\eta_{\bf k}$ is a certain combination of lattice harmonics
($\eta_{{\bf k}=0}=1$). \cite{misha} The first two branches are
soft modes with no dispersion in the nearest-neighbor exchange
approximation acquiring a finite gap above the saturation field
$\tilde{\gamma}H_{c2}=8JS$. Additional interactions produce the
weak dispersion resulting in a finite gaps at ${\bf k}=0$. We
suggest that the lines E and E1-E2 of the spectrum demonstrating
exact linear field dependencies with small gaps at $H=H_{c2}$
($\nu_{1,2}(H_{c2}) = 6$ GHz for $H\parallel [111]$ and 7 and 15
GHz for $H\perp [111]$) correspond to this type of modes. Other
two modes are dispersive magnon branches with the gaps
$\nu_{3,{\bf k}=0}=0$ and $\nu_{4,{\bf k}=0}=\tilde{\gamma}
H_{c2}$. The line D on Fig.~\ref{fig2}-\ref{fig4}) should be
ascribed to the mode $\nu_3$ while the last mode $\nu_4$ lies in a
frequency range of about 200~GHz unreachable in the present
experimental conditions.

At intermediate fields $H_{c1}<H<H_{c2}$ we have observed one
intensive resonance mode which appears above $H_{c1}$ and softens
towards $H_{c2}$ pointing at a continuous second-order nature of
this transition.

In conclusion, the detailed study of the magnetic resonance
properties of a pyrochlore antiferromagnet \GTO\ was carried out.
In the disordered state the system demonstrates a paramagnetic
resonance absorption gradually converting into the nonresonant
absorption of a relaxation type in the vicinity of the ordering
transition. Below the transition, an antiferromagnetic resonance
typical for the nonplanar magnetic ordering was observed. Its low
field part is almost independent on the magnetic field orientation
and qualitatively described in the "hydrodynamic" approximation
with an isotropic tensor of susceptibility. The observed ESR
spectrum is consistent with the $4k$-structure recently proposed
on the basis of a diffusive neutron scattering measurements. In
the spin saturated phase, our experiment provides the direct
observation of the quasi-local soft modes in a pyrochlore magnet.
We also succeeded in tracing their field evolution in magnetic
field above saturation which confirms the theoretical predictions
and previous observations of the magnetocaloric effect.

The authors thank V. Marchenko, O. Petrenko and  V. Glazkov for
valuable discussions. This work was in part supported by INTAS YSF
2004-83-3053, Russian Fund for Basic Researches, Grant 04-02-17294
and RF President Program. S.S.S. is also grateful to National
Science Support Foundation.


\begin{thebibliography}{99}

\bibitem{villain}
J. Villain, Z. Phys. B \textbf{33}, 31 (1979).

\bibitem{ramirez}
A. P. Ramirez, B. S. Shastry, A. Hayashi, {\it et al.}, Phys. Rev.
Lett. \textbf{89}, 067202 (2002).

\bibitem{petrenko}

O. A. Petrenko, M. R. Lees, G. Balakrishnan, D. McK. Paul, Phys.
Rev. B \textbf{70}, 012402 (2004).

\bibitem{bonville}
P. Bonville, J.~A. Hodges, M. Ocio, {\it et al.}, J. Phys.
Condens. Matter. \textbf{15}, 7777 (2003).

\bibitem{stewart}
J.~R. Stewart, G. Ehlers, A.~S. Wills, S.~T. Bramwell and J.~S.
Gardner, J. Phys.: Condens. Matter \textbf{16}, L321 (2004).

\bibitem{raju}
N.~P. Raju, M. Dion, M.~J.~P. Gingras, T.~E. Mason, J.~E. Greedan,
Phys. Rev. B \textbf{59}, 14489 (1999).

\bibitem{cepas}
O. C\'epas, A. P. Young and B. S. Shastry, Phys. Rev. B
\textbf{72}, 184408 (2005).

\bibitem{wills}
A. S. Wills, M. E. Zhitomirsky, B. Canals, {\it et al.}, J. Phys.:
Condens. Matter \textbf{18}, L37 (2006).

\bibitem{palmer}
S. E. Palmer and J. T. Chalker, Phys. Rev. B \textbf{62}, 488
(2000).

\bibitem{lee}
S.-H. Lee, C. Broholm, W. Ratcliff, G. Gasparovic, Q. Huang, T. H.
Kim, and S.-W. Cheong, Nature \textbf{418}, 856 (2002).

\bibitem{misha}
M. E. Zhitomirsky, Phys. Rev. B {\bf 67}, 104421 (2003).

\bibitem{sosin}

S. S. Sosin, L. A. Prozorova, A. I. Smirnov, {\it et al}., Phys.
Rev. B \textbf{71}, 094413 (2005).

\bibitem{oleg}
G. Balakrishnan, O. A. Petrenko, M. R. Lees, and D. McK. Paul, J.
Phys. Condens. Matter \textbf{10}, L723 (1998).

\bibitem{hassan}
A. K. Hassan, L. P. Levy, C. Darie, and P. Strobel Phys. Rev. B
\textbf{67}, 214432 (2003).

\bibitem{marchenko}
A. F. Andreev and V. I. Marchenko, Sov. Phys. Usp. {\bf 23}(1), 21
(1980).

\bibitem{glazkov}
V. N. Glazkov, M. E. Zhitomirsky, A. I. Smirnov, {\it et al.},
Phys. Rev. B {\bf 72}, 020409(R) (2005).

\bibitem{LL}
L. D. Landau and E. M. Lifshits, {\it Quantum Mechanics}
(Pergamon, Oxford, 1980).

\bibitem{Narumi} Y. Narumi, private communication (2005).

\bibitem{yaouanc}
A. Yaouanc, P. D. de Reotier, V. Glazkov, {\it et al}., Phys. Rev.
Lett. \textbf{95}, 047203 (2005).

\end{thebibliography}
\end{document}